# Observation of the Type-II Weyl Semimetal Phase in MoTe$_2$

J. Jiang[1,2,3,4*], Z. K. Liu[1*], Y. Sun[5*], H. F. Yang[2,6], R. Rajamathi[5], Y. P. Qi[5], L. X. Yang[7], C. Chen[2], H. Peng[2], C.-C. Hwang[4], S. Z. Sun[8], S.-K. Mo[3], I. Vobornik[9], J. Fujii[9], S. S. P. Parkin[10], C. Felser[5], B. H. Yan[5], and Y. L. Chen[1,2,7,8]

[1]*School of Physical Science and Technology, ShanghaiTech University, Shanghai, P. R. China*
[2]*Department of Physics, University of Oxford, Oxford, OX1 3PU, UK*
[3]*Advanced Light Source, Lawrence Berkeley National Laboratory, Berkeley, CA 94720, USA*
[4]*Pohang Accelerator Laboratory, POSTECH, Pohang 790-784, Korea*
[5]*Max Planck Institute for Chemical Physics of Solids, D-01187 Dresden, Germany*
[6]*State Key Laboratory of Functional Materials for Informatics, SIMIT, Chinese Academy of Sciences, Shanghai 200050, P. R. China*
[7]*State Key Laboratory of Low Dimensional Quantum Physics, Department of Physics, Tsinghua University, Beijing 100084, P. R. China*
[8]*Hefei Science Center, CAS and SCGY, University of Science and Technology of China, Hefei, P. R. China*
[9]*Istituto Officina dei Materiali (IOM)-CNR, Laboratorio TASC, Trieste, Italy*
[10]*Max Planck Institute of Microstructure Physics, Halle, Germany*

*These authors contributed equally to this work.

**Topological Weyl semimetal (TWS), a new state of quantum matter, has sparked enormous research interest recently. Possessing unique Weyl fermions in the bulk and Fermi arcs on the surface[1-6], TWSs offer a rare platform for realizing many exotic physical phenomena[7-14]. TWSs can be classified into type-I that respect Lorentz symmetry[15-19] and type-II that do not[20-22]. Here, we directly visualize the electronic structure of MoTe$_2$, a recently proposed type-II TWS. Using angle-resolved photoemission spectroscopy (ARPES), we unravel the unique surface Fermi arcs, in good agreement with our *ab-initio* calculations. From spin-resolved ARPES measurements, we demonstrate the non-degenerate spin-texture of surface Fermi-arcs, thereby proving their non-trivial topological nature. Our work not only lead to new understandings of the unusual properties[23] discovered in this family of compounds, but also allow for the further exploration of exotic properties and practical applications of type-II TWSs, as well as the interplay between superconductivity (MoTe$_2$ was discovered to be superconducting recently[24]) and their topological order.**

Three dimensional (3D) topological Weyl semimetals (TWSs) are novel topological quantum materials discovered and intensively investigated recently due to its intimate link between concepts of different fields of physics and material science, as well as the broad application potential[1-6]. In a TWS, low-energy electronic excitations form composite Weyl fermions dispersing linearly along all the three momentum directions across the Weyl points (WPs)[15,16,19,25], which always appear in pairs with opposite chirality. Between WPs of different chirality, there exist intriguing surface Fermi arcs, the unconventional open curve-like Fermi-surfaces (FSs) with non-degenerate spin texture. The exotic bulk and surface electronic structures thus provide an ideal platform for many novel physical phenomena, such as negative magnetoresistance, anomalous quantum Hall effect and chiral magnetic effects[7-11,14,26,27].

Interestingly, the TWSs can be further classified into two types by Fermiology and whether the Lorentz symmetry is respected: type-I TWS that hosts point-like bulk FSs formed solely by WPs (Fig. 1a(i)), which approximately respect the Lorentz symmetry[1-3]; and type-II TWS that breaks the Lorentz symmetry and harbours finite electron density of states at the Fermi-energy (Fig. 1a(ii)). Recently, type-I TWSs have been discovered in the (Ta, Nb)(As, P) family of compounds[16-19,25]. However, the complex 3D crystal structure and the large number (12 pairs) of WPs in these first generation TWSs may pose difficulties in the exploration of novel physical phenomena and practical applications.

More recently, type-II TWS has been proposed to exist in layered transition metal dichalcogenides (TMDs, e.g. $WTe_2$, $MoTe_2$ and $W_xMo_{1-x}Te_2$) with only four pairs of WPs[20-22] (see Fig. 1b, c). With less Weyl points and non-vanishing electronic density of states at the Fermi-surface, type-II TWSs in TMD compounds are expected to show very different properties from the type-I TWSs, such as anisotropic chiral anomaly depending on the current directions, a novel anomalous Hall effect[20]. In addition, the layered nature of the TMD TWSs greatly facilitates the fabrication of devices[28], making them an ideal platform for the realization of novel TWS applications.

In this work, using high resolution angle-resolved photoemission spectroscopy (ARPES), we systematically studied the electronic structure of the orthorhombic ($T_d$) $MoTe_2$. By carrying out broad range photon energy dependent measurements, we can conclude that the observed surface Fermi arcs are in consistent with previous[21] and our *ab-initio* calculations. Moreover, by spin-resolved ARPES measurement, we observed the non-degenerate spin texture of the surface Fermi arcs, further supporting their non-trivial topological nature. The discovery of the TWS phase in MoTe$_2$ could also help understand the puzzling physical properties in the orthorhombic phase of TMDs recently observed[23], and provide a more feasible material platform for the future applications of TWSs due to their layered structures. Furthermore, with the recent discovery of superconductivity in MoTe$_2$[24], it even provides an ideal platform for the study of interplays between superconductivity and the non-trivial topological order.

The crystal structure of T$_d$-MoTe$_2$ is illustrated in Fig. 1d, clearly showing an alternating stacking of Te-Mo-Te triple layer structure (space group *P$_{mn21}$*, with lattice constant *a=6.335Å, b=3.477Å, c=13.883Å* [21]). We have synthesized high quality single crystals for this work: the large flat shinning surface (Fig. 1e, left inset) after in-situ cleaving is ideal for ARPES measurements; and the crystal symmetry is verified by the Laue (before cleaving) and LEED (after cleaving) measurements (Fig. 1e, right inset). The core level photoemission spectrum clearly shows the characteristic Te$_{4d}$ and Mo$_{4p/4s}$ peaks (Fig. 1e), and the broad Fermi surface mapping in Fig. 1f covering multiple Brillouin zones (BZs) confirmed the (001) cleave plane with correct lattice constant. In Fig. 1f, one can already see that the shape of the FS is in general agreements with our calculation (Fig. 1c).

To investigate the detailed electronic structures, we focus on one BZ by carrying out high resolution ARPES measurements, and the results are demonstrated in Fig. 2. In Fig. 2a, the semi-metallic nature of T$_d$-MoTe$_2$ can be clearly seen, where large hole pockets are centered around $\bar{\Gamma}$ and electron pockets are away from $\bar{\Gamma}$ along the $\bar{\Gamma}\bar{Y}$ direction. The evolution of these pockets with binding energy is illustrated in the dispersion plots (Fig. 2b) and the bands' constant energy contours (Fig. 2c). In Fig. 2b, the strongly anisotropic band dispersion along the $\bar{\Gamma}\bar{X}$ and $\bar{\Gamma}\bar{Y}$ directions are clearly shown: the hole- and electron-like Fermi crossings are both observed along $\bar{\Gamma}\bar{Y}$ direction, while only hole-like Fermi crossings can be observed along

$\bar{\Gamma}\bar{X}$, which show broad agreement with our *ab-initio* calculations (Fig. 2b, bottom row). The electron- and hole-pockets' evolution can be better seen in Fig. 2c, where the outer electron pockets shrink with $E_B$ and disappear beyond 0.075 eV, in consistent with the dispersion in Fig. 2b and also our *ab-initio* calculations (see Fig. 2c, bottom row, and comparisons at higher binding energies can be found in Supplementary information). The evolution of the constant energy contours through a larger energy scale is illustrated in Fig. 2d, showing a rich evolution of texture from multiple bands.

In order to distinguish the surface and bulk contributions from numerous band dispersions in Fig. 2, we conducted photon energy dependent ARPES experiment[29] with a broad photon energy range of 20~120 eV, covering more than 5 BZs along the $k_z$ direction (see Fig. 3a, and more details can be found in Supplementary information). In Fig. 3a, throughout the whole range, bands with negligible $k_z$ dispersion can be observed, showing a clear sign of surface states. By plotting the FSs measured with different photon energies (Fig. 3b), one can clearly see that the non-dispersive bands along $k_z$ in Fig. 3a correspond to the FS features in the narrow region between the bulk and electron pockets (as highlighted by the red rectangles in Fig. 3b), which show identical shape under different photon energies (panel i-viii), while the pockets from the bulk states vary strongly (as also can be seen by the dispersive bands in Fig. 3a). This observation is again in good agreement with the *ab-initio* calculation, as shown in Fig. 3c, where the sharp surface state bands clearly emerge between the electron and hole pockets.

After establishing the overall band structures and the identification of surface state regions, we further zoom in to these regions to search for the characteristic surface Fermi arcs, a hallmark of the electronic structure of a TWS. According to the *ab-initio* calculations, although the Weyl points lie slightly above the Fermi level (about 6 and 59 meV above the Fermi-level according to the calculations[21], see Supplementary information for details), the surface Fermi-arcs are still visible, lying within the narrow region between the electron and hole pockets (see Fig. 4a, marked as SA between the points P1 and P2). As there is another surface state (marked as SS) nearby, to best visualize and separate the two surface states, we chose a low (27eV) photon energy to increase the momentum resolution; also under this photon energy, the bulk states intensity is greatly suppressed thus the surface states can be easily seen. Indeed, from these fine measurements, the two surface states in the calculations (Fig. 4a) are both observed (see Fig. 4b, and the measurements from more photon energies can be found in the Supplementary information). Interestingly, as the positions of P1 and P2 vary with binding energy (see Fig. 4a), the surface Fermi-arcs observed in our measurements (Fig. 4b) also vary accordingly (in shape and length), in good agreement with the calculation.

Similar to other topological surface states[29-31], the surface Fermi-arcs should also possess non-degenerate spin texture[1,20,21,32]. Thus to further demonstrate their non-trivial topological nature, we carried out spin-resolved photoemission experiments. The measurements were conducted at the FS and the dashed line in Fig. 4c indicates the sampling direction that cuts through the Fermi arcs (along $k_x$ direction at $k_y=0.24$ Å$^{-1}$). The resulting polarized scattering intensity (Fig. 4d(i,ii)) and the spin polarization asymmetry (Fig. 4d(iii, iv) parallel to $k_x$ and $k_y$ directions are shown in Fig. 4d, clearly indicating the non-degenerate spin texture. Remarkably, while the spin polarization parallel to $k_x$ direction (i.e. $S_x$, see Fig. 4(iii)) switches sign for arcs with positive and negative $k_x$ values, the polarization parallel to $k_y$ direction (i.e. $S_y$, see Fig. 4(iv)) of both arcs have the same sign (though with less magnitude than $S_x$). Such observation of the spin-texture in the surface Fermi-arcs not only demonstrate their non-trivial topology, but also help the design and realization of spin-related physical phenomena and applications.

Our systematic study on the electronic band structures and the observation of surface Fermi-arc states and its spin-texture, together with the broad agreement with the *ab-initio* calculations, establish that $T_d$-MoTe$_2$ is a type-II TWS, which can not only help understand the puzzling physical properties in the orthorhombic phase of TMDs, but also provide a new platform for the realization of exotic physical phenomena and possible future applications.

We note that while we were finalizing this manuscript, three other groups also independently studied this family of compounds and the surface states and arcs (refs 33-35).

**References and notes:**

1 Wan, X., Turner, A. M., Vishwanath, A. & Savrasov, S. Y. Topological semimetal and Fermi-arc surface states in the electronic structure of pyrochlore iridates. *Physical Review B* **83**, 205101 (2011).

2 Weng, H., Fang, C., Fang, Z., Bernevig, B. A. & Dai, X. Weyl Semimetal Phase in Noncentrosymmetric Transition-Metal Monophosphides. *Physical Review X* **5**, 011029 (2015).

3 Bulmash, D., Liu, C.-X. & Qi, X.-L. Prediction of a Weyl semimetal in $Hg_{1-x-y}Cd_xMn_yTe$. *Physical Review B* **89**, 081106 (2014).

4 Burkov, A. A. & Balents, L. Weyl semimetal in a topological insulator multilayer. *Physical Review Letters* **107**, 127205 (2011).

5 Liu, J. & Vanderbilt, D. Weyl semimetals from noncentrosymmetric topological insulators. *Physical Review B* **90**, 155316 (2014).

6 Singh, B. *et al.* Topological electronic structure and Weyl semimetal in the $TlBiSe_2$ class of semiconductors. *Physical Review B* **86**, 115208 (2012).

7 Ashby, P. E. C. & Carbotte, J. P. Magneto-optical conductivity of Weyl semimetals. *Physical Review B* **87**, 245131 (2013).

8 Hosur, P. Friedel oscillations due to Fermi arcs in Weyl semimetals. *Physical Review B* **86**, 195102 (2012).

9 Landsteiner, K. Anomalous transport of Weyl fermions in Weyl semimetals. *Physical Review B* **89**, 075124 (2014).

10 Liu, C.-X., Ye, P. & Qi, X.-L. Chiral gauge field and axial anomaly in a Weyl semimetal. *Physical Review B* **87**, 235306 (2013).

11 Son, D. T. & Spivak, B. Z. Chiral anomaly and classical negative magnetoresistance of Weyl metals. *Physical Review B* **88**, 104412 (2013).

12 Wei, H., Chao, S. P. & Aji, V. Excitonic phases from Weyl semimetals. *Physical Review Letters* **109**, 196403 (2012).

13  Xu, G., Weng, H., Wang, Z., Dai, X. & Fang, Z. Chern semimetal and the quantized anomalous Hall effect in $HgCr_2Se_4$. *Physical Review Letters* **107**, 186806 (2011).

14  Potter, A. C., Kimchi, I. & Vishwanath, A. Quantum oscillations from surface Fermi arcs in Weyl and Dirac semimetals. *Nature Communications* **5**, 5161 (2014).

15  Yang, L. X. *et al.* Weyl semimetal phase in the non-centrosymmetric compound TaAs. *Nature Physics* **11**, 728 (2015).

16  Lv, B. Q. *et al.* Experimental Discovery of Weyl Semimetal TaAs. *Physical Review X* **5**, 031013 (2015).

17  Huang, S. M. *et al.* A Weyl Fermion semimetal with surface Fermi arcs in the transition metal monopnictide TaAs class. *Nature Communications* **6**, 7373, (2015).

18  Xu, S.-Y. et al. Discovery of a Weyl fermion semimetal and topological Fermi arcs. *Science* **349**, 613 (2015).

19  Liu, Z. K. *et al.* Evolution of the Fermi surface of Weyl semimetals in the transition metal pnictide family. *Nature Materials* **15**, 27 (2016).

20  Soluyanov, A. A. *et al.* Type-II Weyl semimetals. *Nature* **527**, 495 (2015).

21  Sun, Y., Wu, S.-C., Ali, M. N., Felser, C. & Yan, B. Prediction of Weyl semimetal in orthorhombic $MoTe_2$. *Physical Review B* **92**, 161107 (2015).

22  Chang, T.-R. *et al*. Arc-tunable Weyl Fermion metallic state in $Mo_xW_{1-x}Te_2$. *arXiv*. 1508.06723 (2015).

23  Qi, Y. *et al.* Superconductivity in Weyl semimetal candidate $MoTe_2$. *Nature Communications* **7**, 11038 (2016).

24  Ali, M. N. *et al.* Large, non-saturating magnetoresistance in $WTe_2$. *Nature* **514**, 205 (2014).

25  Xu, S.-Y. *et al.* Discovery of a Weyl fermion state with Fermi arcs in niobium arsenide. *Nature Physics* **11**, 748 (2015).

26  Huang, X. *et al.* Observation of the Chiral-Anomaly-Induced Negative Magnetoresistance in 3D Weyl Semimetal TaAs. *Physical Review X* **5**, 031023 (2015).

27 Shekhar, C. *et al.* Extremely large magnetoresistance and ultrahigh mobility in the topological Weyl semimetal candidate NbP. *Nature Physics* **11**, 645 (2015).

28 Radisavljevic, B., Radenovic, A., Brivio, J., Giacometti, V. & Kis, A. Single-layer $MoS_2$ transistors. *Nature Nanotechnology* **6**, 147-150 (2011).

29 Chen, Y. Studies on the electronic structures of three-dimensional topological insulators by angle resolved photoemission spectroscopy. *Frontiers of Physics* **7**, 175 (2011).

30 Hasan, M. Z. & Kane, C. L. Colloquium: Topological insulators. *Reviews of Modern Physics* **82**, 3045 (2010).

31 Qi, X.-L. & Zhang, S.-C. Topological insulators and superconductors. *Reviews of Modern Physics* **83**, 1057 (2011).

32 Sun, Y., Wu, S.-C. & Yan, B. Topological surface states and Fermi arcs of the noncentrosymmetric Weyl semimetals TaAs, TaP, NbAs, and NbP. *Physical Review B* **92**, 115428 (2015).

33 Huang, L. *et al.* Spectroscopic evidence fot type II Weyl semimetal state in MoTe2. *arXiv:* 1603.06482 (2016).

34 Deng, K. *et al.* Experimental observation of topological Fermi arcs in type-II Weyl semimetal $MoTe_2$. *arXiv:* 1603.08508 (2016).

35 Belopolski, I. *et al.* Unoccupied electronic structure and signatures of topological Fermi arcs in the Weyl semimetal candidate $Mo_xW_{1-x}Te_2$. *arXiv:* 1512.09099 (2015).

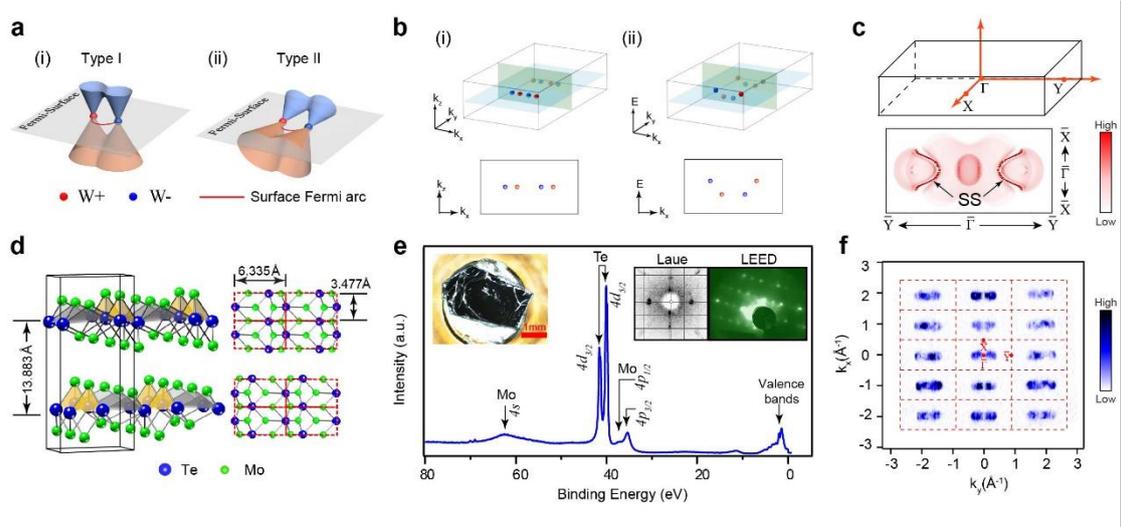

FIG. 1. (color online). **Basic characteristics of type-II TWS and characterization of MoTe$_2$ single crystals. a,** Schematic illustration of type-I and type-II Weyl fermions and Weyl points in the momentum space. **b,** Schematic showing the Weyl points of MoTe$_2$: (i) Upper panel, Weyl points in a BZ; lower panel, the projection to the $k_x$-$k_z$ plane. (ii) Upper panel, Weyl points in the energy-momentum space; lower panel, the projection to the $k_x$-$E$ plane. **c,** Top: Illustration of the BZ of T$_d$-MoTe$_2$ with high symmetry point marked. Bottom: Fermi surface by our *ab-initio* calculation, showing the bulk electron, hole pockets and the surface states (marked as "SS"). **d,** Crystal structure of the T$_d$-MoTe$_2$, showing alternating "…A-B-A-B…" stacking of MoTe$_2$ layers. **e,** Core level spectrum, showing characteristic Mo$_{4s}$, Te$_{4p/4d}$ core level peaks. Left inset: cleaved surface of MoTe$_2$ single crystal used in this study. Right insets: Laue and LEED pattern showing the T$_d$ phase of MoTe$_2$. **f,** Broad range photoemission spectral intensity map of the Fermi-surface covers more than 10 BZ's, showing the correct symmetry and the characteristic FS.

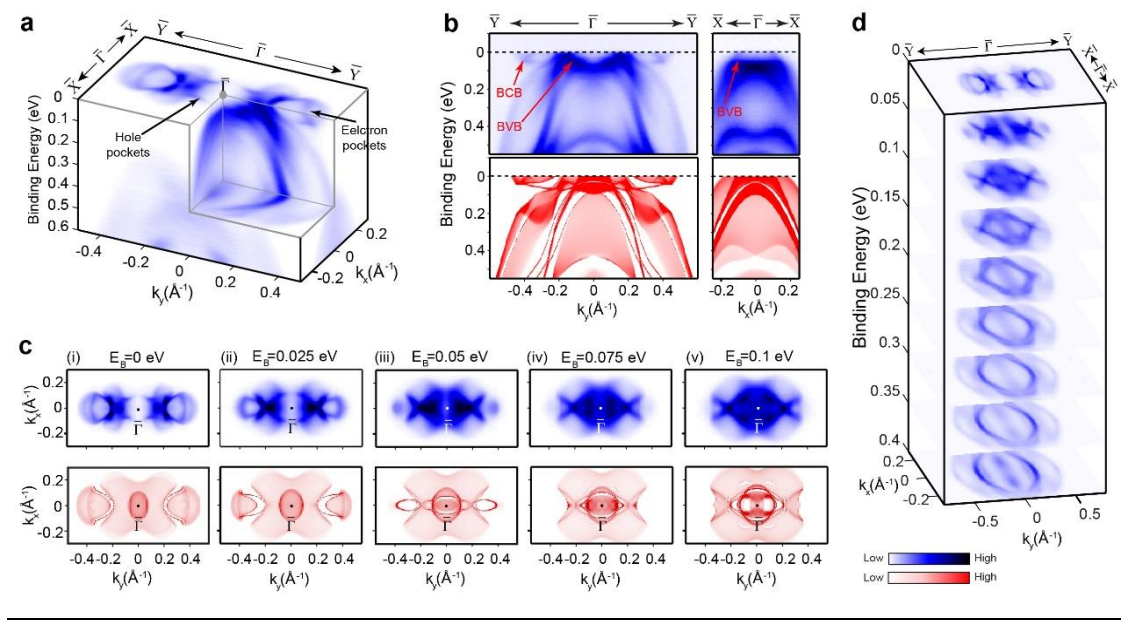

FIG. 2. (color online). **General electronic structure of $T_d$-MoTe$_2$. a,** 3D intensity plot of the photoemission spectra around $\bar{\Gamma}$, with the electron and hole pockets indicated. **b**, Top row: High symmetry cut along the $\bar{Y}\bar{\Gamma}\bar{Y}$ and $\bar{X}\bar{\Gamma}\bar{X}$ directions, respectively. "BCB" and "BVB" stand for "bulk conduction band" and "bulk valance band", respectively. Bottom row: The corresponding calculations along the $\bar{Y}\bar{\Gamma}\bar{Y}$ and $\bar{X}\bar{\Gamma}\bar{X}$ directions, respectively. **c**, Top row: photoemission spectral intensity map showing the constant energy contours of bands at $E_B$=0, 0.025 eV, 0.05 eV, 0.075 eV and 0.1 eV, respectively. Bottom row: Corresponding calculated constant energy contours at the same binding energies as in experiments above. **d**, Stacking plots of constant-energy contours in broader binding energy range show the band structure evolution.

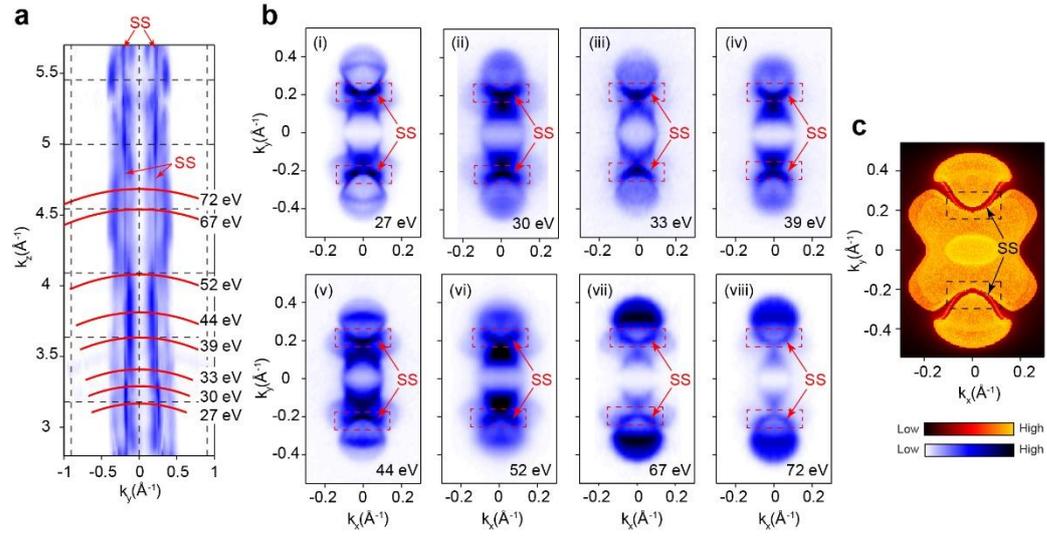

FIG. 3. (color online). **Bulk and surface electronic states of MoTe$_2$ probed by photon energy dependent ARPES measurements. a,** The photoemission spectral intensity map in $k_y$-$k_z$ plane, where the red arrows indicate the surface states (marked as "SS") that show no dispersion along the $k_z$ direction. Red curves indicate the $k_z$ locations of photon energies used for the measurements in **b**. **b,** Fermi surfaces measured under different photon energies as indicated in **a,** Red dashed rectangles and arrows highlight the position of the surface states which does not change shape with the photon energies. **c,** Calculated Fermi surface with black dashed rectangles highlight the same area as in **b,** indicating clear surface states inside.

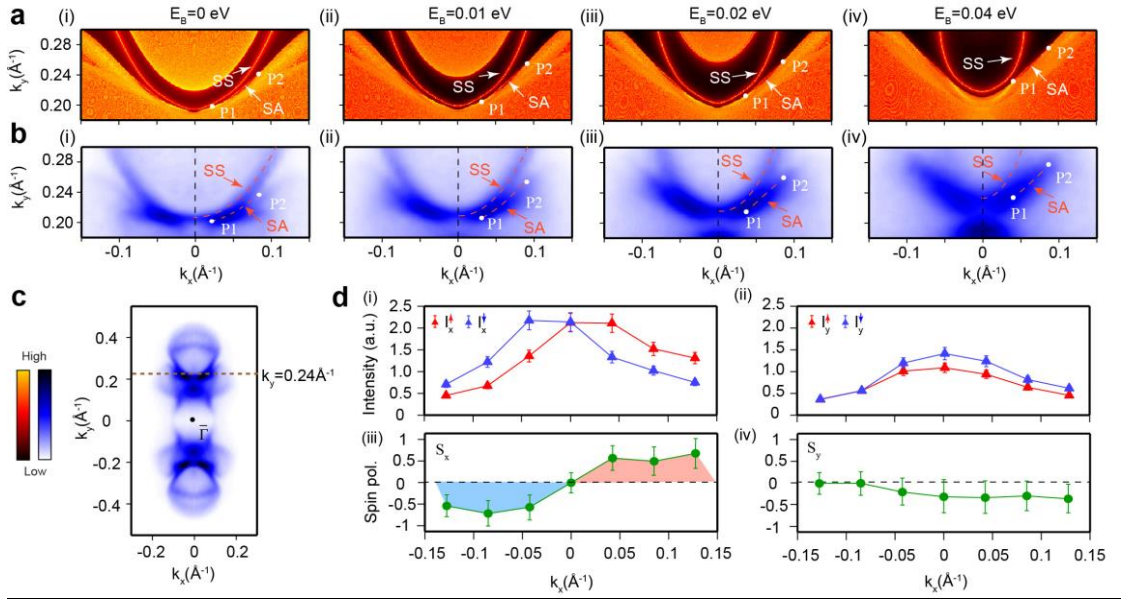

FIG. 4. (color online). **Evolution of the Fermi arcs with binding energies and their spin-texture. a,** Evolution of the projection of the Fermi arcs in the calculation at $E_B$=0, 0.01 eV, 0.02 eV and 0.04 eV, respectively. **b,** The corresponding photoemission intensity map in the same area as in **a**. The position of SS is extracted from the spectra and the positions of the P1 and P2 are extracted from the calculation (i.e. identical to those in **a**). **c,** The Fermi surface of the sample measured for spin-resolved ARPES, where the dashed line indicates the momentum positions where the spin-resolved measurements were performed. **d,** (i, ii) scattering intensity from two scattering targets orthogonal to each other. Red and blue symbols represent the scattering intensity with each scattering target when magnetized to up/down polarization, respectively. (iii, iv), the asymmetry calculated along the $k_x$ and $k_y$ directions from (i, ii) by $S_{x/y} = \frac{I_{x/y}^{Up} - I_{x/y}^{down}}{I_{x/y}^{Up} + I_{x/y}^{down}}/\eta$, respectively (where $\eta \approx 0.3$ is the Sherman function of the spin-detector), showing clearly the switching of sign for $S_x$ (iii) while not for $S_y$ (iv).